# Observation of Correlation between Route to Formation, Coherence, Noise, and Communication Performance of Kerr Combs

Pei-Hsun Wang,[1,*] Fahmida Ferdous,[1] Houxun Miao,[2,3] Jian Wang,[1,4] Daniel E. Leaird,[1] Kartik Srinivasan,[2] Lei Chen,[2] Vladimir Aksyuk,[2] and Andrew M. Weiner[1,4,*]

[1]*School of Electrical and Computer Engineering, Purdue University, 465 Northwestern Avenue, West Lafayette, IN 47907-2035*
[2]*Center for Nanoscale Science and Technology, National Institute of Standards and Technology, 100 Bureau Dr, Gaithersburg, MD 20899, USA*
[3]*Nanocenter, University of Maryland, College Park, MD 20742, USA*
[4] *Birck Nanotechnology Center, Purdue University, 1205 West State Street, West Lafayette, Indiana 47907, USA*
*[\*wang1173@purdue.edu](mailto:wang1173@purdue.edu), [amw@purdue.edu](mailto:amw@purdue.edu)*

**Abstract:** Microresonator optical frequency combs based on cascaded four-wave mixing are potentially attractive as a multi-wavelength source for on-chip optical communications. In this paper we compare time domain coherence, radio-frequency (RF) intensity noise, and individual line optical communications performance for combs generated from two different silicon nitride microresonators. The comb generated by one microresonator forms directly with lines spaced by a single free spectral range (FSR) and exhibits high coherence, low noise, and excellent 10 Gbit/s optical communications results. The comb generated by the second microresonator forms initially with multiple FSR line spacing, with additional lines later filling to reach single FSR spacing. This comb exhibits degraded coherence, increased intensity noise, and severely degraded communications performance. This study is to our knowledge the first to simultaneously investigate and observe a correlation between the route to comb formation, the coherence, noise, and optical communications performance of a Kerr comb.

## 1. Introduction

Generation of optical frequency combs via cascaded four-wave mixing (FWM) in high quality factor (Q) microresonators has received considerable attention [1]. Tuning a continuous-wave laser into a cavity resonance leads to build-up of intracavity power, resulting in parametric gain that enables additional cavity modes to oscillate. Frequency combs have been demonstrated in a variety of materials, including silica [2-6], calcium fluoride [7-9], magnesium fluoride [10-12], fused quartz [13], and silicon nitride [14-18]. The recent demonstration [5, 14] of comb generation from planar microresonators that can be fabricated on-chip using standard microlithography tools offers additional prospects for robust implementations that may be suitable for applications such as multi-wavelength optical communications.

   Initial studies concentrated on characterizing the properties of the generated optical spectra, including the optical bandwidth and the uniformity of the frequency spacing. More recently, increasing attention has been devoted to other properties of the generated combs. For example, noise has been characterized both in the vicinity of radio-frequency (RF) tones generated through beating of comb lines on a photodetector [13, 17, 19] and at baseband [18]. Pulse shaping experiments have provided insight into the time domain behavior and especially the coherence of the generated combs [13, 15, 20]. A few experiments have looked at optical communications performance using individual lines selected from a microresonator comb as the light source [21, 22]. Although in some cases error-free communications has been observed, in other cases communications performance is badly degraded. It is now becoming important to correlate the information provided by these different measurement modalities and to establish how to select or design microresonator devices for desired operation.

In previous time domain studies in our laboratory, in which we performed pulse shaping, compression, and autocorrelation measurements on combs from planar silicon nitride microresonators, we identified two distinct routes to comb formation [15, 20]. In one case, which we termed Type I, the FWM mixing cascade proceeds from initial sidebands spaced one free spectral range (FSR) from the pump. In a second case, the initial pair of sidebands is generated with $N$ FSR spacing from the pump, where $N$ is an integer greater than one. Cascaded FWM then results in a comb with $N$ FSR frequency spacing. In both cases the cascaded FWM process is expected to produce frequency spacings that are precisely equal across the comb, resulting in high temporal coherence as from a mode-locked laser. Consistent with this expectation, our experiments provided evidence of high coherence by demonstrating high quality compression into bandwidth-limited pulses. However, in the latter case, tuning the pump closer into resonance gives rise to independent FWM processes that originate from the initial, multiple-FSR-spaced lines. As a result, additional comb lines are generated that eventually yield a comb with single FSR spacing, a route to comb formation which we termed Type II. Such Type II combs involve an imperfect frequency division process which leads to degraded compression and reduced coherence in pulse shaping experiments [15, 20]. An important point is that these experiments establish a link between the route to comb formation and the coherence of the generated comb.

Other studies provide evidence of connections between different comb properties and routes to comb formation. For example, ref. [13] reported distinct regimes of operation, in which one regime exhibited high coherence in pulse shaping studies and high spectral purity in RF spectrum analyzer measurements of tones generated through beating of comb lines, while another regime exhibited both low optical coherence and degraded RF spectra. In another experiment RF beating measurements obtained via multi-heterodyne mixing with a self-referenced comb from a mode-locked laser were used to chart the evolution of Type II combs obtained from silicon nitride and magnesium fluoride microresonators [17]. The data show that individual comb lines exhibit spectral substructure too fine to be resolved in normal optical spectrum analyzer measurements and link such spectral substructure to increased noise. In the current paper we investigate temporal coherence (as revealed through pulse shaping & compression), intensity noise, and optical communications performance of combs generated from two different silicon nitride microresonators. A Type I comb, generated from one microresonator, exhibits high coherence, low noise, and excellent 10 Gbit/s optical communications results. A Type II comb, generated from the second microresonator, displays degraded coherence, increased intensity noise, and severely degraded communications performance. To the best of our knowledge, these results constitute the first observation of a simultaneous correlation between the route to comb formation and the coherence, noise, and optical communications performance of a Kerr comb.

## 2. Experimental Setup

Figure 1 shows a schematic diagram of the experimental setup. As in our previous studies [15, 20], a tunable continuous-wave (CW) laser is amplified and launched into a ring resonator fabricated in silicon nitride, which results in generation of multiple frequency components (a frequency comb) through cascaded FWM. Two resonators are tested, both with rings with 40 μm outer radius. One resonator, which we refer to as Resonator 1, has 2 μm × 430 nm waveguide cross-section and loaded quality factor ($Q$) of $6\times10^5$. This resonator generates a Type I comb, characterized by lines generated at single free spectral range (FSR) spacing and high coherence. The second resonator, which we refer to as

Resonator 2, has 2 µm × 550 nm waveguide cross-section and loaded $Q$ of $2\times10^6$. The normalized transmission at resonance is ≈ 0.2 for both resonators. Initially this resonator generates lines spaced by two FSRs. Upon further tuning into resonance, new frequencies are generated, resulting in a spectrum with lines spaced by single FSR. This forms a Type II comb, which has reduced coherence.

A key goal of our study is to compare coherence and intensity noise behavior of Type I and Type II combs. A programmable pulse shaper plays an important role in each set of experiments. For intensity noise and communications measurements, the pulse shaper simply selects an individual line out of the generated comb, blocking the others. The power of the selected line is boosted by an erbium-doped fiber amplifier (EDFA), which then passes through a tunable filter with 0.4 nm bandwidth to suppress amplified spontaneous emission (ASE) noise. The output is detected by a photodiode with 12 GHz bandwidth, after which the low frequency intensity noise is characterized by an RF spectrum analyzer. On-off keying (OOK) communications experiments are performed by modulating the selected comb line at 9.953 Gbit/s using a lithium niobate intensity modulator driven by a length $2^{31}$-1 non-return-to-zero pseudorandom bit sequence (PRBS). The photodetected output is connected to a bit error rate tester (BERT) for error analysis and a sampling oscilloscope with bandwidth 60 GHz for display of eye diagrams. Eye diagrams and bit error rate data are reported only for filtered comb lines with optical signal to noise ratio (OSNR)≧15 dB (based on 0.1 nm OSA spectral resolution). This ensures that any degradation of the communications results is associated with the inherent properties of selected comb lines themselves as opposed to poor OSNR.

To provide information on the coherence, we follow the procedure we reported in [15], in which the pulse shaper manipulates the spectral phase of the generated comb on a line-by-line basis in an attempt to achieve pulse compression. In these experiments the pulse shaper may also attenuate the pump line, which often remains very strong, to a level comparable to the neighboring lines. The output of the pulse shaper is amplified and connected to an intensity autocorrelator based on second harmonic generation in a noncollinear, background-free geometry. The dispersion of the fibers connecting the microresonator chip and the autocorrelator are left uncompensated in these experiments. The pulse shaper varies the spectral phases one at a time in order to maximize the second harmonic signal at zero delay. Full autocorrelations are then recorded as a function of delay. Experimental autocorrelations are compared with those calculated on the basis of the measured comb spectrum with the assumption of flat spectral phase. High quality pulse compression showing good agreement between experimental and calculated autocorrelation traces provides evidence of stable spectral phase and good coherence [15, 20].

**3. Results and Discussions**

Figure 2(a) shows the optical spectrum obtained from Resonator 1. The comb spectra are measured directly after the resonator. The CW laser is set at 1543.08 nm with 26.5 dBm input power. The average line spacing is around 4.79 nm. For Resonator 2, the initial sidebands were first formed with 2 FSR spacing at 13 dBm pump power (as shown in Fig. 2(b)). By red-shifting the pumping wavelength around 0.01 nm (≈1 GHz), additional comb lines filled in and formed a Type II comb spectrum with one FSR spacing (Fig. 2(c)) at the same pumping power. The average spacing for the Type II combs is around 4.74 nm.

Figures 3(a), (b), and (c) show the autocorrelation traces after phase compensation by the pulse shaper (blue traces) for combs corresponding respectively to Figs. 2(a), (b), and (c). The red lines are the intensity autocorrelation traces calculated based on the measured spectra after the pulse shaper and assuming flat spectral phase, while the green and blue traces show the measured autocorrelation traces before and after the line-by-line phase compensation. For the comb from Resonator 1 and also the comb with 2 FSR spacing from Resonator 2, the measured autocorrelation traces after pulse compression are in close agreement with those calculated. This demonstrates successful pulse compression and a high level of coherence. For the Type II comb from Resonator 2 with pumping at 1549.94 nm, the on-off contrast of the experimental autocorrelation (Fig. 3(c)) is significantly degraded compared to that of the simulated traces. In general, the reduction in autocorrelation contrast ratio indicates reduced coherence as well as degraded compression [23, 24]. These results are consistent with those reported previously by our group [15, 20]. The uncertainty of the calculated curves is determined by repeating the simulations for 3 spectra recorded sequentially during the autocorrelation measurement. The largest variation in the calculated curves is observed for the case of Fig. 3(c) at a delay equal to half of the autocorrelation period, where the minimum and maximum calculated values differ by only 2.3 % of the mean calculated value. This difference is almost invisible by eye on the scale of the figure and is much smaller than the difference in experimental and calculated on-off contrast.

We next present noise measurements. Figure 4 shows intensity noise spectra of individual comb lines selected out of the generated combs over the range 0 MHz to 700 MHz, measured using an RF spectrum analyzer with 1.5 MHz resolution bandwidth. For the Type I comb from Resonator 1, the RF spectra for three different selected lines are essentially identical and are at the noise floor of the spectrum analyzer (Fig. 4(a)). Figure 4(b) shows the RF spectra of the 1559.4 nm comb line generated from Resonator 2 with slightly different pumping wavelengths. When the pump is set at 1549.93 nm, a multiple FSR comb is generated, and a low-noise state RF spectrum is observed, similar to Fig. 4(a). However, when the pump is tuned slightly (0.01 nm) into resonance, a Type II comb is formed, and strong intensity noise peaks are observed in the RF spectrum at 156 MHz ± 0.7 MHz and harmonics. These data illustrate the very rapid onset of intensity noise with the transition to Type II operation. Similar noise peaks are observed when other single lines of the comb are selected for analysis in the Type II regime (Fig. 4(c)). The presence of these noise peaks may be explained in terms of beating between combs with different offset frequencies that overlap in the same region of optical frequency [17].

We now report the optical communications measurements. Figure 5(a) shows bit error rate (BER) data as a function of the received optical power when individual comb lines produced from Resonator 1 are selected. Results are plotted for each of the seven comb lines within the EDFA bandwidth. The performance for each of the generated comb lines is essentially identical to the reference curve, which is measured for the CW pump laser (1543.08 nm) prior to the microresonator. Figure 5(b) shows the corresponding eye diagrams at -11 dBm received power. The clean and open eyes again indicate the suitability of the Type I comb as a multi-wavelength source for high quality communications. At -11 dBm received power, each of the selected comb lines exhibited error-free operation (no errors are observed over 300 s measurement time, corresponding to BERs below $10^{-12}$).

We now discuss the communications measurements from comb lines obtained from Resonator 2. For the initial, 2-FSR-spaced comb, high quality communications is again observed for each of the three comb lines with power high enough to yield good OSNR after amplification. As an example, Fig. 6(a) shows a nicely open eye diagram for the 1559.4 nm comb line; similar high quality eyes are observed for the other high power lines at 1540.6 nm and 1549.9 nm. The corresponding BER traces are shown in Fig. 6(b); in each case high quality communications performance is observed, similar to the results from Resonator 1. Error-free operation could also be achieved.

However, quite different results are obtained when the pump laser is tuned 0.01 nm further into resonance for Type II operation. Now all of the generated comb lines show badly degraded eye diagrams. Figure 7(a) presents the eye diagram obtained for 1559.4 nm comb line. Although this is the same line as presented in Fig. 6(a) for the multiple-FSR-spaced comb, and the power (-11 dBm) is similar, now the eye is nearly half closed, with a high level of intensity noise clearly observed on the top rail of the eye, corresponding to the 'on' state. Figure 7(b) shows the eye diagram of the 1554.6 nm line, which is one of the lines that first appears after the transition to Type II operation. Now the eye is completely closed. This is in qualitative agreement with an observable broadening of the peaks in Fig. 4(c) compared to those in Fig. 4(b), which corresponds to an increase in the integrated intensity noise. These finding suggest that in the absence of a transition to a low-noise state [21], Type II combs are unsuitable for optical communications.

**4. Conclusion**

We have investigated combs generated from two different silicon nitride microresonators with identical 40 μm radius. In each case we assess the time domain coherence of the full comb (through pulse shaping and autocorrelation experiments) and measure intensity noise and communications performance (through ≈10 Gbit/s on-off keying experiments) of individual lines selected from the comb. For the first resonator, the comb forms directly with lines spaced by a single FSR (termed a Type I comb). For the second resonator, the comb initially forms with multiple FSR line spacing; additional lines later fill in to reach single FSR spacing (the latter is termed a Type II comb). Both the Type I comb from the first resonator and initial multiple-FSR-spaced comb from the second resonator exhibit high coherence, low intensity noise, and excellent 10 Gbit/s optical communications results. In contrast, the Type II comb exhibits reduced coherence, increased intensity noise, and severely degraded optical communications performance. This study is to our knowledge the first to simultaneously investigate and observe a correlation between the route to comb formation, the coherence, noise, and optical communications performance of a Kerr comb. Our observations should prove useful in selection of microresonator devices providing combs suitable for optical communications applications.

This work was supported in part by the National Science Foundation under grant ECCS-1102110 and by the Air Force Office of Scientific Research under grant FA9550-12-1-0236. Dr. Houxun Miao acknowledges support under the Cooperative Research Agreement between the University of Maryland and the National Institute of Standards and Technology Center for Nanoscale Science and Technology, Award 70NANB10H193, through the University of Maryland.

**References**
1. T. J. Kippenberg, R. Holzwarth, and S. A. Diddams, "Microresonator-based optical frequency combs," Science **332**, 555–559 (2011).
2. P. Del'Haye, A. Schliesser, O. Arcizet, T. Wilken, R. Holzwarth, and T. J. Kippenberg, "Optical frequency comb generation from a monolithic microresonator," Nature **450**, 1214-1217 (2007).
3. I. H. Agha, Y. Okawachi, M. A. Foster, J. E. Sharping, and A. L. Gaeta, "Four-wave-mixing parametric oscillations in dispersion-compensated high-Q silica microspheres," Phys. Rev. A **76**, 043837 (2007).
4. P. Del'Haye, O. Arcizet, A. Schliesser, R. Holzwarth, and T. J. Kippenberg, "Full stabilization of a microresonator-based optical frequency comb," Phys. Rev. Lett. **101**, 053903 (2008).
5. L. Razzari, D. Duchesne, M. Ferrera, R. Morandotti, S. Chu, B. E. Little, and D. J. Moss, "CMOS-compatible integrated optical hyper-parametric oscillator," Nat. Photonics **4**, 41-45 (2010).
6. P. Del'Haye, T. Herr, E. Gavartin, R. Holzwarth, and T. J. Kippenberg, "Octave spanning tunable frequency comb from a microresonator," Phys. Rev. Lett. **107**, 063901 (2011).
7. A. A. Savchenkov, A. B. Matsko, V. S. Ilchenko, I. Solomatine, D. Seidel, and L.Maleki, "Tunable optical frequency comb with a crystalline whispering gallery mode resonator," Phys. Rev. Lett. **101**, 093902 (2008).
8. I. S. Grudinin, N. Yu, and L. Maleki, "Generation of optical frequency combs with a $CaF_2$ resonator," Opt. Lett. **34**, 878-880 (2009).
9. A. A. Savchenkov, A. B. Matsko, W. Liang, V. S. Ilchenko, D. Seidel, and L. Maleki, "Kerr combs with selectable central frequency," Nat. Photonics **5**, 293-296 (2011).
10. W. Liang, A. A. Savchenkov, A. B. Matsko, V. S. Ilchenko, D. Seidel, and L. Maleki, " Generation of near-infrared frequency combs from a $MgF_2$ whispering gallery mode resonator," Opt. Lett. **36**, 2290-2292 (2011).
11. C. Y. Wang, T. Herr, P. Del'Haye, A. Schliesser, J. Hofer, R. Holzwarth, T. W. Hänsch, N. Picqué, and T. J. Kippenberg, "Mid-infrared optical frequency combs based on crystalline microresonators," arXiv:1109.2716v2.

12. I. S. Grudinin, L. Baumgartel, and N. Yu, "Frequency comb from a microresonator with engineered spectrum," Opt. Express **20**, 6604-6609 (2012).
13. S. B. Papp and S. A. Diddams, "Spectral and temporal characterization of a fused-quartz-microresonator optical frequency comb," Phy. Rev. A **84**, 053833 (2011).
14. J. S. Levy, A. Gondarenko, M. A. Foster, A. C. Turner-Foster, A. L. Gaeta, and M. Lipson, "CMOS-compatible multiple-wavelength oscillator for on-chip optical interconnects," Nat. Photonics **4**, 37-40 (2010).
15. F. Ferdous, H. Miao, D. E. Leaird, K. Srinivasan, J. Wang, L. Chen, L. T. Varghese, and A. M. Weiner, "Spectral line-by-line pulse shaping of on-chip microresonator frequency combs," Nat. Photonics **5**, 770-776 (2011).
16. M. A. Foster, J. S. Levy, O. Kuzucu, K. Saha, M. Lipson, and A. L. Gaeta, "Silicon-based monolithic optical frequency comb source," Opt. Express **19**, 14233-14239 (2011).
17. T. Herr, K. Hartinger, J. Riemensberger, C. Y. Wang, E. Gavartin, R. Holzwarth, M. L. Gorodetsky, and T. J. Kippenberg, "Universal formation dynamics and noise of Kerr-frequency combs in microresonators," Nat. Photonics **6**, 480-487 (2012).
18. Y. Okawachi, K. Saha, J. S. Levy, Y. H. Wen, M. Lipson, and A. L. Gaeta, "Octave-spanning frequency comb generation in a silicon nitride chip," Opt. Lett. **36**, 3398-3400 (2011).
19. J. Li, H. Lee, T. Chen, and K. J. Vahala, "Low-pump-power, low-phase-noise, and microwave to millimeter-wave repetition rate operation in microcombs," arXiv:1208.5256v1.
20. F. Ferdous, H. Miao, P.-H. Wang, D. E. Leaird, K. Srinivasan, L. Chen, V. Aksyuk, and A. M. Weiner, " Probing coherence in microcavity frequency combs via optical pulse shaping," Opt. Express **20**, 21033–21043 (2012).
21. J. S. Levy, K. Saha, Y. Okawachi, M. A. Foster, A. L. Gaeta, and M. Lipson, "High-performance silicon-nitride-based multiple-wavelength source," IEEE Photon. Technol. Lett. **24**, 1375-1377 (2012).
22. J. Pfeifle, C. Weimann, F. Bach, J. Riemensberger, K. Hartinger, D. Hillerkuss, M. Jordan, R. Holtzwarth, T. J. Kippenberg, J. Leuthold, W. Freude, and C. Koos, "Microresonator-based optical frequency combs for high-bitrate WDM data transmission," in *Optical Fiber Communication Conference*, OSA Technical Digest (Optical Society of America, 2012), paper OW1C.4.
23. E. P. Ippen and C. V. Shank, "Techniques for measurement," in *Ultrashort Light Pulses: Picosecond Techniques and Applications*, S. L. Shapiro, ed. (Springer-Verlag, 1977) vol. **18**, pp. 83–122.
24. A. M. Weiner, *Ultrafast Optics* (Wiley, 2009), Chap. 3.

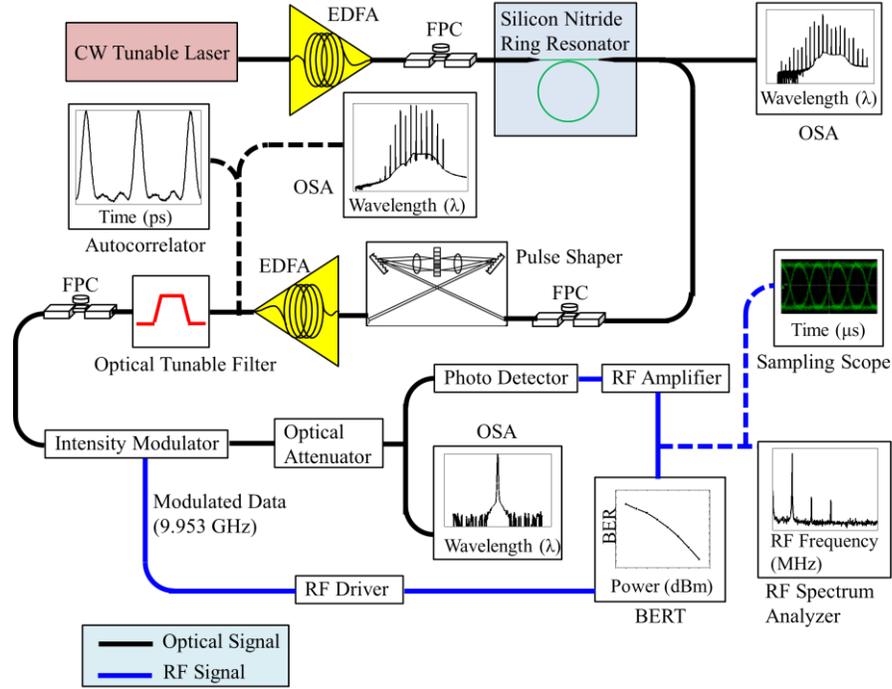

Fig. 1. Schematic diagram of the experimental setup for optical communications measurement of the frequency combs generated from silicon nitride microresonators. CW: continuous-wave; RF: radio-frequency; EDFA: erbium doped fiber amplifier; FPC: fiber polarization controller; OSA: optical spectrum analyzer; BERT: bit error rate tester.

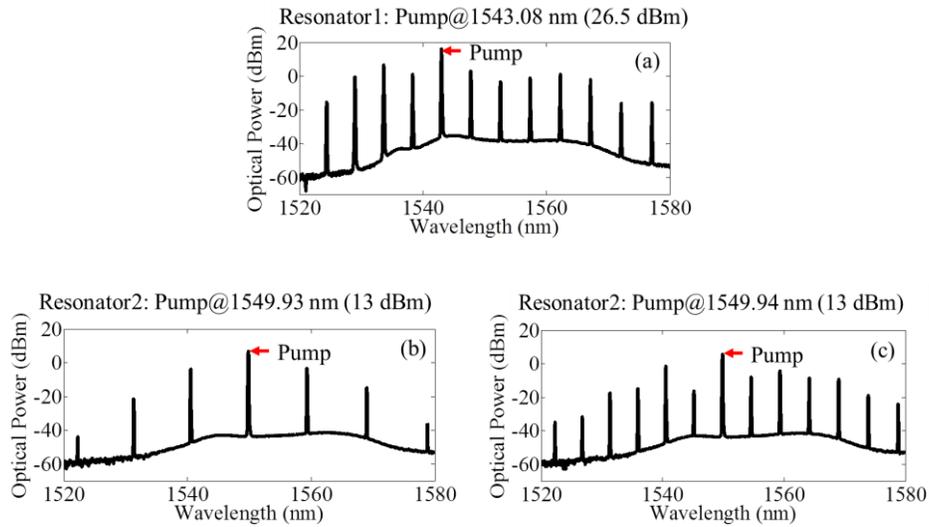

Fig. 2. Generated frequency combs from two different silicon nitride ring resonators (outer radius, 40 μm). The spectral curves are obtained under different regimes of comb formation. The input power and wavelength of the tunable CW pump are (a) 26.5 dBm, 1543.08 nm (Resonator 1), (b) 13 dBm, 1549.93 nm (Resonator 2), and (c) 13 dBm, 1549.94 nm (Resonator 2), respectively. OSA spectra are measured directly after the microresonator chip at 0.01 nm spectral resolution.

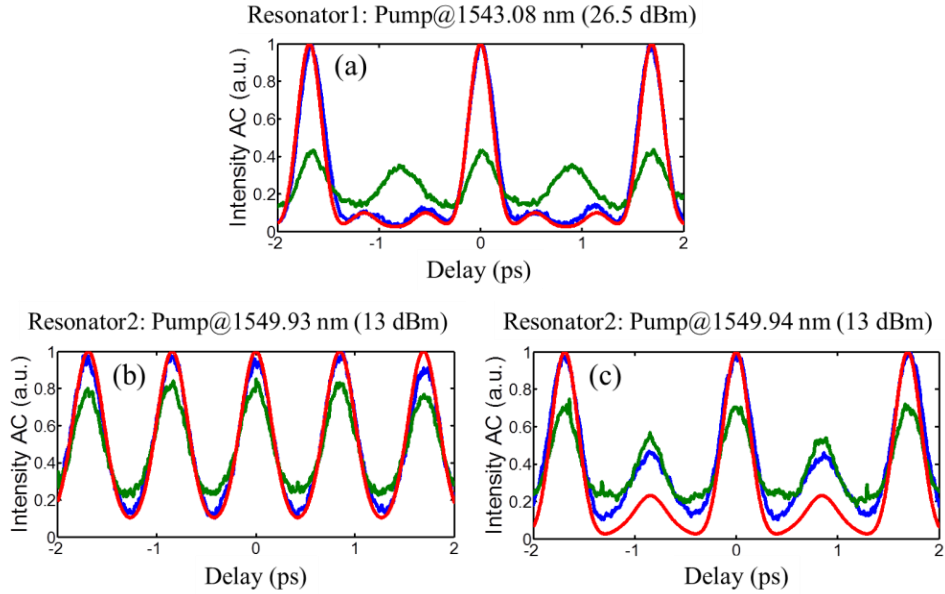

Figs. 3. (a) (b) (c) Autocorrelation traces corresponding to the frequency combs in Figs. 2(a), (b), and (c), respectively. Red lines show intensity autocorrelation traces calculated by taking the optical spectra and assuming flat spectral phase, while the green and blue traces show the measured autocorrelation traces before and after the line-by-line phase compensation.

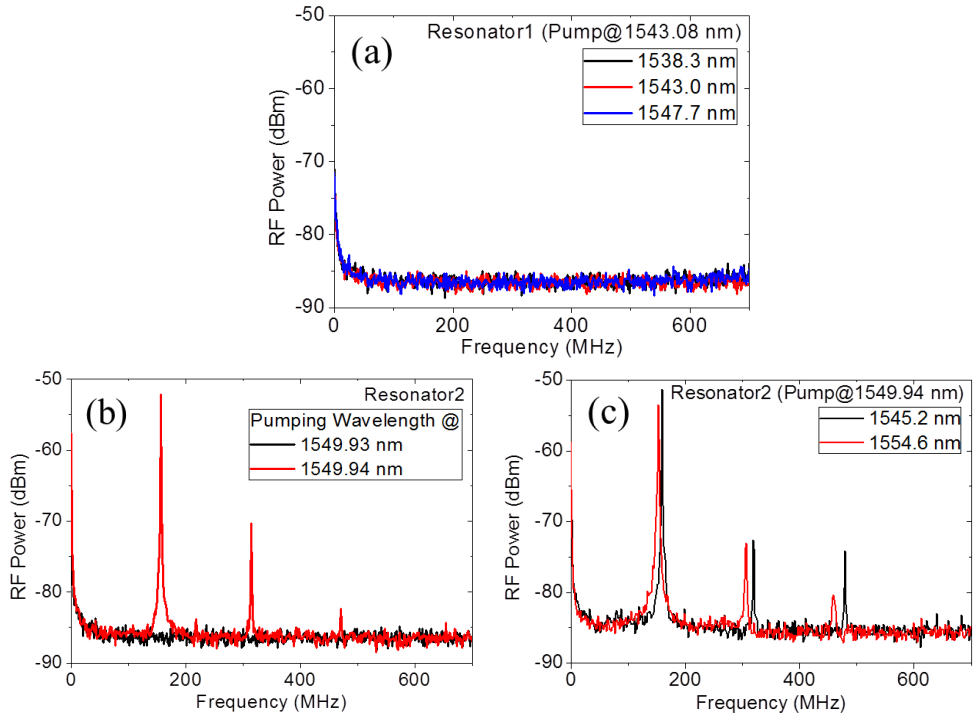

Fig. 4. (a) RF noise spectra of exemplary combs obtained from Resonator 1 (Type I comb). (b) RF noise spectra of the frequency comb at 1559.4 nm before and after tuning the pumping wavelength (Resonator 2). (c) The corresponding RF spectra of the newly generated spectral lines from a Type II frequency comb with pumping wavelength at 1549.94 nm (Resonator 2).

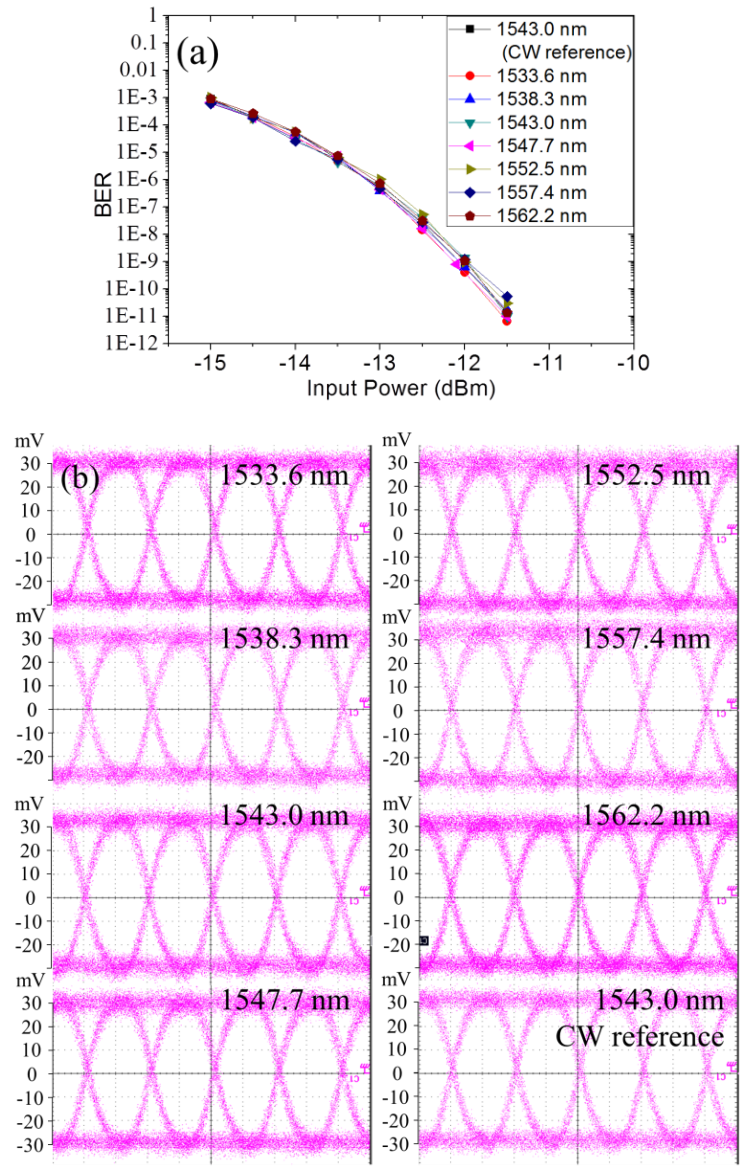

Fig. 5. (a) BER measurements and (b) the corresponding eye diagrams of seven selected comb lines from Type I combs and the CW reference (taken prior to the microresonator). The open eye diagrams confirm the high quality of the OOK communications performance.

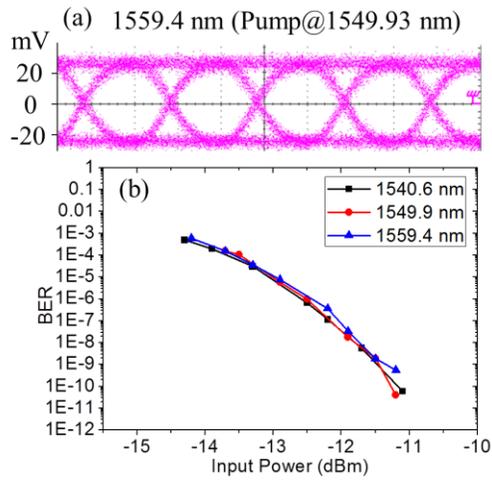

Fig. 6. Representative eye diagrams and BER measurements of the filtered comb lines generated from Resonator 2. The received power of the plotted eye diagrams is -11 dBm. (a) The eye diagram of the 1559.4 nm comb line and (b) BER traces of individual lines selected from Resonator 2 with pumping at 1549.93 nm. High quality communications is again observed for these three comb lines.

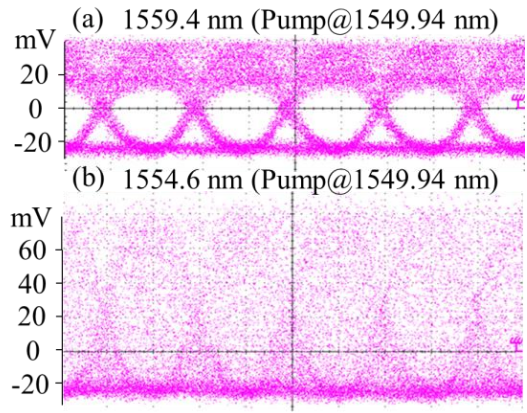

Fig. 7. The eye diagrams of the (a) 1559.4 nm and (b) 1554.6 nm comb line from a Type II frequency comb with pumping at 1549.94 nm.